\documentclass[10pt, a4paper]{article}
\usepackage[T1]{fontenc}
\usepackage[
	a4paper,
	left=1.4in,right=1.4in,
	top=1.4in,bottom=1.6in,
	heightrounded
]{geometry}

\usepackage{lmodern}
\usepackage[final]{microtype}

\usepackage{mathtools}
\usepackage{physics}
\usepackage{amsfonts, amssymb}
\usepackage{graphicx}

\usepackage{booktabs}

\usepackage{titlesec}
\titleformat{\section}
{\bfseries\large}
{\thesection}{0.8em}{}

\titleformat{\subsection}
{\bfseries\normalsize}
{\thesubsection}{0.8em}{}

\titleformat{\subsubsection}
{\bfseries\normalsize}
{\thesubsubsection}{0.8em}{}
\usepackage{caption}
\DeclareCaptionLabelFormat{scap}{[#1 #2]}
\usepackage{hyperref}
\usepackage[sorting=none]{biblatex}
\graphicspath{{/}}
\usepackage{multirow}
\usepackage{authblk}

\title{Flexible numerical characteristic gluing to extremal Reissner--Nordström black holes}
\author{{Junwoo Lee\thanks{jwl52@cam.ac.uk}}}
\affil{\small{\emph{Department of Applied Mathematics and Theoretical Physics, University of Cambridge,} \\ \emph{Wilberforce Road, Cambridge CB3 0WA, United Kingdom}}}
\date{\small{\today}}

\begin{document}
\maketitle
\thispagestyle{empty}
\begin{abstract}
	Recent work has used the method of characteristic gluing to construct spacetimes violating the third law of black hole mechanics. This paper uses numerical methods to investigate two aspects of the gluing construction of solutions describing the gravitational collapse of a charged scalar field to form an extremal Reissner--Nordström black hole.
	First, for a massless scalar field, we improve significantly the lower bounds on the size of the final black hole for which gluing is possible. Second, for a massive scalar field, it is known that the formation of an extremal Reissner--Nordström black hole is impossible if the mass-to-charge ratio of the scalar field exceeds unity.
	We present strong evidence that this bound is sharp by constructing gluing solutions with a scalar field mass-to-charge ratio just below unity.
\end{abstract}
\vspace{2pt}
\section{Introduction}
The \emph{third law of black hole mechanics} was originally stated by Bardeen, Carter and Hawking~\cite{bardeen1973four}, and states that the surface gravity of a stationary black hole cannot reach zero by finite operations; it was reformulated by Israel~\cite{israel1986third} in a more dynamical form that an extremal black hole cannot be formed from a non-extremal black hole in a finite time under regularity/energy assumptions.
However, this was proven to be false by Kehle and Unger~\cite{kehle2022gravitational}, who constructed counterexamples in the Einstein-Maxwell massless charged scalar field system in spherical symmetry. Their proof was based on the ``characteristic gluing'' method, first developed by Aretakis, Czimek and Rodnianski~\cite{aretakis2025characteristic,aretakis2024linear,aretakis2023kerr,czimek2022obstruction}.

The simplest example of Kehle and Unger involves specifying data on a null hypersurface that, at early times, agrees exactly with the data on an outgoing null hypersurface in Minkowski spacetime, and, at late times, agrees exactly with the data on the event horizon of an extremal Reissner--Nordström (eRN) black hole. The free datum in the gluing construction is the complex scalar field in the region of the null hypersurface between the Minkowski and eRN regions. Characteristic initial/final value problems are then solved to construct a solution to the future and past of the null hypersurface. Kehle and Unger proved that the free datum can be chosen to produce solutions with \(C^k\) regularity across the null hypersurface, for any desired value of \(k\). These solutions describe gravitational collapse to form an exact eRN black hole in finite time.

A limitation of the proof is that it works only for \emph{sufficiently large} black holes: they require \(eM \gg 1\) where \(e>0\) is the scalar field charge coupling and \(M\) is the mass of the final eRN black hole. It is interesting to ask whether such solutions also exist when \(eM\) is not large. This problem is a suitable candidate for numerical exploration by constructing the gluing solution explicitly. This was studied by Gadioux, Reall and Santos~\cite{gadioux2025formation}, who showed that a sufficient condition for existence of solutions of \(C^k\) gluing is \(eM > 9.27,\,78.5,\,4417\) for \(k=0,\,1,\,2\), respectively.\footnote{\(k=0\) gluing solutions might not be expected to be physical but the results for \(k=0\) share many similarities with the results for higher \(k\) \cite{gadioux2025formation}.} Very recently, Schneider has proved that a necessary condition for gluing (for any \(k\)) is \(eM>1/3\) and that a sufficient condition for \(k=0\) gluing is \(eM > 1.707\)~\cite{schneider2026chargebound}.

The first goal of this paper is to use numerical methods to significantly improve the sufficient condition on \(eM\) under which gluing is possible for various \(k\). The ansätze considered in~\cite{gadioux2025formation} were low-dimensional, with the number of free parameters matching the number of gluing conditions required for a given \(k\). This is a rather restrictive assumption. More recently, Crump, Gadioux, Reall and Santos~\cite{crump2026violation} considered characteristic gluing in five-dimensional vacuum gravity and numerically constructed a solution describing finite-time formation of extremal rotating black holes. In that work, a more flexible ansatz motivated by a neural network architecture was employed. In this ansatz, the number of free parameters is much greater than the number of equations to be solved.

In this paper, we will revisit the Minkowski-to-eRN gluing problem numerically using the more flexible neural network ansatz of~\cite{crump2026violation} and more specialised numerical methods. We will find that this enables us to significantly improve the lower bounds on \(eM\) above which gluing is possible. We will also obtain results for \(k=3\), which is the minimum regularity necessary to guarantee that, to the past of the null hypersurface, the spacetime has \(C^2\) regularity at the origin~\cite{kehle2022gravitational}.

It is also possible to include a mass parameter \(m\) for the scalar field. Reall has proved that it is impossible to form an eRN black hole in gravitational collapse if \(m/e \geq 1\)~\cite{reall2024third}. (A third law also holds in this case~\cite{mcsharry2025supersymmetric}.)
However, it is not known whether this bound is sharp. This was studied numerically by Gadioux, Reall and Santos in~\cite{gadioux2025formation}, who constructed gluing solutions with \(m/e=0.2773,0.1103,0.00992\) for \(k=0,1,2\) respectively. In this paper we will present strong numerical evidence that the bound of~\cite{reall2024third} is indeed sharp by numerically constructing gluing solutions for \(k=1,2,3\) with \(m/e\) just below, but very close to, 1.

A summary of the main results is given in Table~\ref{tab:comparison-lambda-zero}.
This paper is organised as follows. In Section~\ref{sec:gluing_problem}, we introduce the gluing equations to be solved and decompose the complex scalar field into its amplitude and phase. In Section~\ref{sec:numerical_implementation}, we explain in detail the numerical setup and the algorithms used. Then we present the results of the numerical gluing in Section~\ref{sec:results}, starting with the \(C^0\) gluing, followed by massless higher-regularity gluing, and finally the gluing with a massive scalar field. We summarise the results and discuss future directions in Section~\ref{sec:discussion}.

\noindent\textit{Note Added.} During the preparation of this work, I asked OpenAI's GPT-6 Astra whether the analytical lower bound of \( eM > 1 /3 \)~\cite{schneider2026chargebound} could be improved, and it generated an argument for the stronger bound \( eM > 1 /2 \), which has not yet been checked. Meanwhile, Hod and Piran independently came up with a quantum-mechanical argument leading them to conjecture that \( eM\geq 1 /2 \)~\cite{hod2026third}.
\section{Characteristic Gluing Problem}\label{sec:gluing_problem}
\subsection{The Reissner--Nordström Black Hole}\label{sec:RN}
In a spacetime with zero cosmological constant, the Reissner--Nordström metric is (in units \(G=1\))
\[ds^2=-fdt^2+f^{-1}dr^2+r^2d\Omega_2^2\]
where
\begin{equation*}
	f(r)=1-\frac{2M}{r}+\frac{Q^2}{r^2}
	\label{eq:RN_metric}
\end{equation*}
and \(d\Omega_2^2\) is the metric of the unit 2-sphere.
Its horizon radii are given by \(r_\pm=M\pm\sqrt{M^2-Q^2}\), so that extremality is achieved when \(M=|Q|\) and the horizon radius is \(r_+=M\).
\subsection{Equations of Motion and Characteristic Gluing}\label{sec:eom}
We follow the Einstein-Maxwell charged Klein-Gordon equations and the formulation from~\cite{kehle2022gravitational} and~\cite{gadioux2025formation}. The spherically symmetric metric is given by
\[ds^2=-\Omega^2(U,V)dUdV+r^2(U,V)d\Omega_2^2\]
where \(U,V\) are double-null coordinates, \(\Omega\) is a conformal factor and \(r\) is the area radius.
Let \(\Phi=\Phi(U,V)\) be the complex scalar field, and pick a gauge such that the electromagnetic potential is given by \(A=A_U(U,V)dU\) and the covariant derivative is given by \(D_U\Phi=\partial_U\Phi+i e A_U\Phi\).
Then under spherical symmetry, the electromagnetic tensor becomes
\[F=\frac{Q\Omega^2}{2r^2}dU\wedge dV\]
so that \(Q=Q(U,V)\) becomes the charge enclosed by the 2-sphere with specified \(U\) and \(V\).
Following~\cite{kehle2022gravitational}, the \emph{renormalised Hawking mass} \(\varpi\) is defined quasilocally by\footnote{Henceforth, \(U,\,V\) subscripts denote derivatives, e.g. \(r_U = \partial_U r\).}
\begin{equation}
	1-\frac{2\varpi}r+\frac{Q^2}{r^2}=g^{\alpha\beta}(dr)_\alpha(dr)_\beta=-\frac{4r_Ur_V}{\Omega^2}.
	\label{eq:varpi}
\end{equation}
This definition allows quasilocal classification of a symmetry sphere in non-stationary spacetime: subextremal if \(\varpi>|Q|\), extremal if \(\varpi=|Q|\) and superextremal if \(\varpi<|Q|\)~\cite{gadioux2025formation}.

The equations of motion in the double-null coordinates \((U,V)\) are given as follows.
\begin{itemize}
	\item Raychaudhuri equations:
	      \begin{equation}
		      \partial_V(r_V/\Omega^2)=-\frac{r}{\Omega^2}|\Phi_V|^2,
		      \label{eq:raychaudhuri}
	      \end{equation}
	      \begin{equation}
		      \partial_U(r_U/\Omega^2)=-\frac{r}{\Omega^2}|D_U\Phi|^2.
		      \label{eq:raychaudhuri_U}
	      \end{equation}
	\item Maxwell equations:
	      \begin{equation}
		      Q_V=er^2\Im(\Phi\overline{\Phi_V}),
		      \label{eq:Q_V}
	      \end{equation}
	      \begin{equation}
		      Q_U=-er^2\Im(\Phi\overline{D_U\Phi}),
	      \end{equation}
	      \begin{equation}
		      A_{UV}=-\frac{Q\Omega^2}{2r^2}.
		      \label{eq:A_UV}
	      \end{equation}
	\item Wave equations:
	      \begin{equation}
		      r_{UV}=-\frac{\Omega^2}{4r}-\frac{r_Vr_U}{r}+\frac{\Omega^2Q^2}{4r^3}+\frac{m^2\Omega^2r|\Phi|^2}4,
		      \label{eq:r_UV}
	      \end{equation}
	      \begin{equation}
		      \partial_V \partial_U \log \left(\Omega^2\right)=\frac{\Omega^2}{2 r^2}+\frac{2 r_Ur_V}{r^2}-\frac{\Omega^2 Q^2}{r^4}-2 \operatorname{Re}\left(D_U \Phi \overline{\Phi_V}\right)
	      \end{equation}
	      and
	      \begin{equation}
		      \Phi_{UV}=-\frac{\Phi_Ur_V+r_U\Phi_V}{r}+\frac{ie\Omega^2Q\Phi}{4r^2}-ie\frac{A_U\Phi r_V}{r}-ieA_U\Phi_V-\frac{m^2\Omega^2\Phi}{4}.
		      \label{eq:Phi_UV}
	      \end{equation}
\end{itemize}
We will take the null hypersurface of the gluing construction to be the surface $U=0$. For $V<0$ we choose the data on this hypersurface to coincide with data on an outgoing null cone of Minkowski spacetime, and for $V>1$, we choose the data to coincide with data on the event horizon of an eRN black hole of mass $M$ (so $r_+ = M=Q$). We will prescribe the complex scalar field \(\Phi(V)\) along the portion \(\mathcal{C}=\{0\leq V\leq 1, U=0\}\) of the null hypersurface, with the gauge choice \(\Omega|_\mathcal{C}\equiv 1\) without loss of generality~\cite{gadioux2025formation}. We need to choose this profile to ensure that when we solve Equations~\eqref{eq:raychaudhuri}-\eqref{eq:Phi_UV} we achieve \(r(1)=r_+=M\), \(r_V(1)=0\), \(Q(1)=M\), \(Q(0)=0\), alongside the zero (renormalised) Hawking mass condition \(\varpi(0)=0\), i.e.\ \(r_U(0)=-1/(4r_V(0))\). 
In particular, for a gluing solution to be of regularity \(C^k\), we require that \( \Phi \in C^k([0,1];\mathbb{C}) \), and \(\partial^j_U\Phi(1)=0\) and \(\partial_V^j\Phi=0\) for \(j=0,1,\ldots,k\) at endpoints \( V=0,\,1 \)~\cite{kehle2022gravitational}.
It must also satisfy the \emph{admissibility conditions} \(\inf_{V\in[0,1]}r(V)>0\) and \(\sup_{V\in[0,1]}r_U(V)<0\), for the well-definedness of the metric along the gluing surface and to exclude the antitrapped surface, respectively~\cite{kehle2022gravitational}.\footnote{Refer to~\cite{kehle2022gravitational,gadioux2025formation} for a more detailed approach to the gluing problem and the spacetime construction.}

Given the scalar field \(\Phi(V)\), the boundary and admissibility conditions for the gluing can be calculated in the following order.
First, Equation~\eqref{eq:raychaudhuri} is solved backwards from \(V=1\) to \(V=0\). Then Equation~\eqref{eq:Q_V} is integrated from \(V=0\) to find \(Q(V)\). Using these, we integrate Equation~\eqref{eq:r_UV} to find \(r_U(V)\) and check the admissibility condition \(\sup_Vr_U(V)<0\). Finally Equations~\eqref{eq:A_UV} and~\eqref{eq:Phi_UV} are integrated to find \(\Phi_U(V)\). Higher-order \(U\)-derivatives can be obtained by differentiating
Equations~\eqref{eq:raychaudhuri}-\eqref{eq:Phi_UV}; the integrated forms are written in Appendix~\ref{sec:higher_order_gluing}. The endpoint conditions for the \(V\)-derivatives are imposed by the form of the scalar field, as we will discuss in Section~\ref{sec:the_ansatz}.
\subsection{Amplitude-Phase Decomposition}\label{sec:amp_phase_decomposition}
We decompose the complex scalar field \(\Phi(V)\) into its amplitude and phase as
\begin{equation}
	\Phi(V)=\rho(V)e^{-i\omega(V)}.
	\label{eq:scalar_field}
\end{equation}
Under this decomposition,
\begin{equation}
	E(V)\coloneq|\Phi_V(V)|^2=\rho_V^2+\rho^2\omega_V^2,\qquad \Im(\Phi\overline{\Phi_V})=\rho^2\omega_V.
	\label{eq:focusing}
\end{equation}
This allows us to rewrite the equations in terms of \(\rho\) and \(\omega\). Write \(r(V)=r_+\xi(V)\) where \( r_+ \) is the horizon radius (Section~\ref{sec:RN}). Then Equation~\eqref{eq:raychaudhuri} can be rewritten as
\begin{equation}
	\xi''(V)=-E(V)\xi(V)
	\label{eq:xi_raychaudhuri}
\end{equation}
with \(\xi(1)=1\) and \(\xi'(1)=0\). We follow the order of calculations displayed in Section~\ref{sec:eom}. Integrating Equation~\eqref{eq:Q_V},
\begin{equation}
	Q(V)=er_+^2\int_0^VdV'\xi^2\rho^2\omega_V.
	\label{eq:Q-integral}
\end{equation}
In particular, for the extremal Reissner--Nordström black hole, we have
\begin{equation}
	I\coloneq\int_0^1\xi^2\rho^2\omega_V dV=\frac 1{eM}.
	\label{eq:I_def}
\end{equation}
Equation~\eqref{eq:r_UV} can be written in the integrated form as
\begin{equation}
	r(V)r_U(V)-r(0)r_U(0)=\int_0^V\left(-\frac{1}{4}+\frac{Q^2}{4r^2}+\frac{m^2r^2\rho^2}{4}\right)dV'.
	\label{eq:rr_U}
\end{equation}

Integrating Equation~\eqref{eq:xi_raychaudhuri} gives \(\xi'(V)=\int_V^1E(V')\xi(V')dV'\). Integrating once more and swapping the order of integration gives
\[1-\xi(0)=\int_0^1V\xi(V)E(V)dV.\]
But observe that if \(\xi\) is positive, then \(\xi'\) is non-negative, so \(\inf_V\xi(V)=\xi(0)\). Hence, the admissibility condition \( \inf_V\xi(V)>0 \) implies the constraint
\begin{equation}
	\int_0^1V\xi(V)E(V)dV<1
	\label{eq:focusing_constraint}
\end{equation}
for a gluing solution to satisfy.
\section{Numerical Implementation}\label{sec:numerical_implementation}
We address two main problems: determining (1) the lowest value of \(eM\) for which the gluing is possible, and (2) the highest value of \(m/e\) for which the gluing is possible.
As \(eM\) decreases (or \(m/e\) increases), the admissibility constraints become more restrictive, as seen in~\cite{gadioux2025formation}, and further explained in Sections~\ref{sec:hrmg} and~\ref{sec:massive_scalar_field}.
Furthermore, in our computations, it was observed that numerical root-finding becomes more difficult in these limits; thus we employ a \emph{continuation algorithm} to gradually follow the solution branches, starting from easier conditions for gluing (e.g.\ larger \(eM\) and/or smaller \(m/e\)).\footnote{The \( C^0 \) gluing is dealt with separately, see Section~\ref{sec:direct_minimisation}.}

To obtain a \( C^{2}  \) spacetime solution for the Einstein-Maxwell charged scalar field system, we require a \(C^3\) gluing solution on the characteristic hypersurface \(\mathcal{C}\), since a derivative is lost at \(r=0\)~\cite{kehle2022gravitational}.
We nevertheless explore \(C^0\), \(C^1\), \(C^2\) and \(C^3\) characteristic gluing since it was observed in~\cite{gadioux2025formation} that lower regularity solutions retain many qualitative features of the higher regularity solutions, while being numerically less demanding; this also allows us to examine how the solutions and the bounds change as additional regularity conditions are imposed.

In this section, we introduce the ansatz used, the continuation method, and the \emph{direct minimisation} approach used specifically for the \(C^0\) gluing.
The numerical details are further explained in Appendix~\ref{sec:numerical_details}.

\subsection{The Ansatz}\label{sec:the_ansatz}
We consider shallow neural-network ansätze with a \(\tanh\) activation function, multiplied by polynomial envelopes, for the amplitude and phase of the scalar field.
Following the notation of Equation~\eqref{eq:scalar_field}, for the \( C^k \) gluing,
\begin{equation}
	\begin{aligned}
		\rho(V)   & =\chi_{i,k,p}(V),         \\
		\omega(V) & =\kappa V+\chi_{1,0,p}(V)
	\end{aligned}
\end{equation}
where \(\chi_{i,k,p}\) is the \emph{neural network ansatz} (or \emph{\(\tanh\) ansatz}) defined as
\begin{equation}
	\chi_{i,k,p}(V)=P_{i,k}(V)\left[a_0+\sum_{j=1}^{p}a_j\tanh(w_jV+b_j)\right].
	\label{eq:tanh-ansatz}
\end{equation}
An additional linear term is included in the phase to better represent the linear growth.

A polynomial envelope \(P_{i,k}(V)\) is introduced to ensure \(C^k\) regularity of the ansatz at the endpoints \(V=0\) and \(V=1\) (Section~\ref{sec:eom}): the scalar field and its first \( k \) \( V \)-derivatives vanish at \( V=0,\,1 \).
In this work, we use two types of polynomial envelopes: \(P_{1,k}=V^{k+1}(1-V)^{k+1}\) (\emph{Type 1}) and
\[P_{2,k}(V)=\begin{cases}
		I_{V/\delta_1}(k+1,k+1)     & 0\leq V<\delta_1              \\
		1                           & \delta_1\leq V\leq 1-\delta_2 \\
		I_{(1-V)/\delta_2}(k+1,k+1) & 1-\delta_2<V\leq 1
	\end{cases}
\]
(\emph{Type 2}) following~\cite{crump2026violation}, where \(I_x(a,b)=B(x;a,b)/B(a,b)\) is the \emph{regularised incomplete beta function}, with \(B(x;a,b)\) and \(B(a,b)\) being the incomplete and complete beta functions, respectively. A distinctive feature of the Type 2 envelope is that it rises and falls from/to zero sharply near the endpoints, allowing a sharp variation of the scalar field near the endpoints.
There are a total of \( 3p+1 \) free parameters in Equation~\eqref{eq:tanh-ansatz}, and we denote them by
\begin{equation}
	\tilde{\theta}=(a_0,a_1,\ldots,a_p,w_1,\ldots,w_p,b_1,\ldots,b_p).
	\label{eq:theta}
\end{equation}
Then the scalar field can be characterised by \( \theta =(\tilde{\theta} _{\text{amp}},\log\kappa ,\tilde{\theta} _{\text{phase}}) \) where \( \tilde{\theta } _\text{amp} \) and \( \tilde{\theta }_{\text{phase}} \) are defined as in Equation~\eqref{eq:theta}, corresponding to the \( \tanh \) parts of the amplitude and phase, respectively (Equation~\eqref{eq:tanh-ansatz}).
\subsection{Residual}\label{sec:residual}
The gluing conditions given in Section~\ref{sec:eom} can be presented in terms of the residual \(F_k(\theta)\), defined by
\begin{equation}
	F_k(\theta)\coloneq\begin{pmatrix}
		Q(1;\theta)-M              \\
		\Re \partial_U\Phi(1;\theta)       \\
		\Im \partial_U\Phi(1;\theta)       \\
		\vdots                     \\
		\Re \partial_U^k\Phi(1;\theta) \\
		\Im \partial_U^k\Phi(1;\theta)
	\end{pmatrix},
	\label{eq:residual}
\end{equation}
for the \(C^k\) gluing, where \(\theta\) is the parameter vector of the ansatz defined in Section~\ref{sec:the_ansatz}.
We say that \(\theta\) is a candidate gluing solution of regularity \(C^k\) if \(F_k(\theta)=0\).
The final black hole mass was normalised to \( M=1 \) in the numerical calculations.
For a candidate solution to be a valid gluing solution, recall that it must also satisfy the admissibility conditions \(\inf_{V\in[0,1]}\xi(V)>0\) and \(\sup_{V\in[0,1]}r_U(V)<0\).
We implement these conditions as \( \inf_V\xi(V)>\delta_\xi \) and \( \sup_V r_U(V)<-\delta_{r_U} \) for some small positive constants \( \delta_\xi \) and \( \delta_{r_U} \) (both set to \( 10^{-4} \) unless stated otherwise) to account for numerical errors.
The residual and admissibility conditions are evaluated numerically by following the chain of calculations described in Sections~\ref{sec:eom} and~\ref{sec:amp_phase_decomposition}.

\subsection{Finding the Gluing Solutions}
\subsubsection{Direct Minimisation}\label{sec:direct_minimisation}
In the case of \(C^0\) gluing, the residual becomes a scalar function \(F_0(\theta)=Q(1;\theta)-M\), i.e.\ for a given \( eM \) we only need to satisfy the charge-matching condition given in Equation~\eqref{eq:I_def}, with \( I \) regarded as a function of the parameter vector, \( I=I(\theta) \).
Hence, minimising \( eM \) corresponds to the (direct) minimisation problem
\[\begin{aligned}
		\min_{\theta\in\mathbb{R}^{6p+3}} & \;-\log I(\theta)                                                           \\
		\text{subject to}                 & \;\inf_V\xi(V;\theta)>\delta_\xi, \quad \sup_V r_U(V;\theta)<-\delta_{r_U},
	\end{aligned}\]
which we solve using the sequential least squares programming (SLSQP) algorithm~\cite{kraft1988software} implemented in the SciPy package~\cite{virtanen2020scipy}.
Each parameter vector obtained from the SLSQP algorithm is validated independently using more accurate numerical schemes (Appendix~\ref{sec:numerical_details}).
We start the minimisation from 100 random parameter vectors satisfying the admissibility condition with \( p=16 \).
The better-performing runs are then continued by increasing \( p \) up to 1024, while the others are discarded. Increasing \(p\) gives the ansatz more freedom to represent the scalar field profile, while using many different initial guesses helps us find lower values of \(eM\).

\subsubsection{Continuation Algorithm}\label{sec:continuation_algorithm}
For regularities higher than \( C^0 \), the residual is vector-valued, and the \(U\)-derivatives must be matched in addition to the charge. Directly minimising \(eM\), as in Section~\ref{sec:direct_minimisation}, is therefore less attractive, since it would require solving multiple nonlinear equations simultaneously in an increasingly restricted parameter space, making the result more sensitive to the initial guess and valid solutions harder to find.\footnote{Direct minimisation was tried for \( C^1 \) gluing, but failed to improve from the initial seed.}

Instead, a natural way to solve a nonlinear system is by \emph{numerical continuation}~\cite{allgower1990numerical}. We start from a seed with large \( eM \) (or low \( m /e \)) and then proceed by small steps, finding solutions with slightly modified \(eM\) and/or \( m /e \).

The parameter vector for the initial seed is sampled randomly (Appendix~\ref{sec:numerical_details}), and it is successively corrected.
That is, for a seed of regularity \( C^k \), we start by solving \( F_0=0 \), then \( F_1=0 \), until we reach \( F_k=0 \). The constrained Levenberg--Marquardt (LM) algorithm~\cite{levenberg1944method, marquardt1963algorithm} is employed: we minimise  \(\|F_k(\theta)\|_2^2/2\) subject to the admissibility conditions \(\inf_V\xi(V)>0\) and \(\sup_V r_U(V)<0\). We correct the parameter vector multiple times on successively finer grids.
This strategy allows us to find a valid solution more easily than by matching all residuals simultaneously, especially for \( C^2 \) and \( C^3 \) gluing. We used both seeds obtained from the randomised discovery procedure described above and structured or heuristic seeds informed by preliminary numerical exploration.

One standard form of a numerical continuation is the ``predictor--corrector'' continuation~\cite{allgower1990numerical}.
The parameter vector for a new step is first predicted by extrapolation; this allows the numerical method to \emph{follow} the solution branch. For the \( eM \)-decrease process, the continuation parameter is \( \lambda \coloneq 1 /eM \). Suppose we have accepted solutions from the two preceding steps: \((\theta_{n-1},\lambda_{n-1})\) and \((\theta_n,\lambda_n)\). Then for a continuation step \(\Delta\lambda\), the candidate continuation parameter is \(\lambda_*=\lambda_n+\Delta\lambda\), and the initial guess \(\theta_*\) for the next step is
\[\theta_*=\theta_n+\frac{\lambda_*-\lambda_n}{\lambda_n-\lambda_{n-1}}(\theta_n-\theta_{n-1}).\]
For the \( m /e \)-increase process, the continuation parameter is \(m/e\). Here, \( eM \) is allowed to vary as well, and the value of \( \log (eM) \) is extrapolated similarly.
The predicted parameter vector is then corrected using the constrained LM algorithm, as in the initial seed discovery.
Once a solution is found, it is validated on finer grids and independent numerical schemes before it is accepted, and the continuation proceeds to the next step.

It is possible for the solution found to satisfy the admissibility conditions by a very small margin. We then take a few steps at the same \( eM \) but with improved admissibility conditions, i.e.\ increase \( \inf_V\xi(V;\theta) \) and/or decrease \( \sup r_U(V;\theta) \); we refer to this as \emph{margin rebuilding}. If the continuation stagnates due to the limited dimension of the parameter vector, then \( p \) is increased provided that it is effective for the next step, up to a maximum of \( p=2048 \).

We note that the method of slowly varying \( eM \) and \( m /e \) allows us to find solutions which are rather challenging to find directly, but it has an apparent limitation that the method is \emph{branch-dependent}, i.e.\ distinct initial seeds lead to distinct branches of solutions.

\section{Results}\label{sec:results}

\subsection[The C0 Gluing]{The \(C^0\) Gluing}\label{sec:C0_res}
The \(C^0\) gluing was studied using the direct minimisation approach described in Section~\ref{sec:direct_minimisation}, with the Type 1 (\(P_{1,0}\)) envelope.
We restrict our attention to the massless (\( m /e =0\)) case for the \( C^0 \) gluing since the mass parameter \( m \) does not enter the residual \( F_0 \) and only affects the admissibility condition \( \sup_Vr_U<0 \), which is fundamentally different from the higher-regularity cases where the mass parameter also affects the residual. 
The massive case is considered in Section~\ref{sec:massive_scalar_field}, where we consider \( C^1 \), \( C^2 \) and \( C^3 \) gluing, which is also a \( C^0 \) solution.

\begin{figure}[!htbp]
	\begin{center}
		\includegraphics[width=0.98\textwidth]{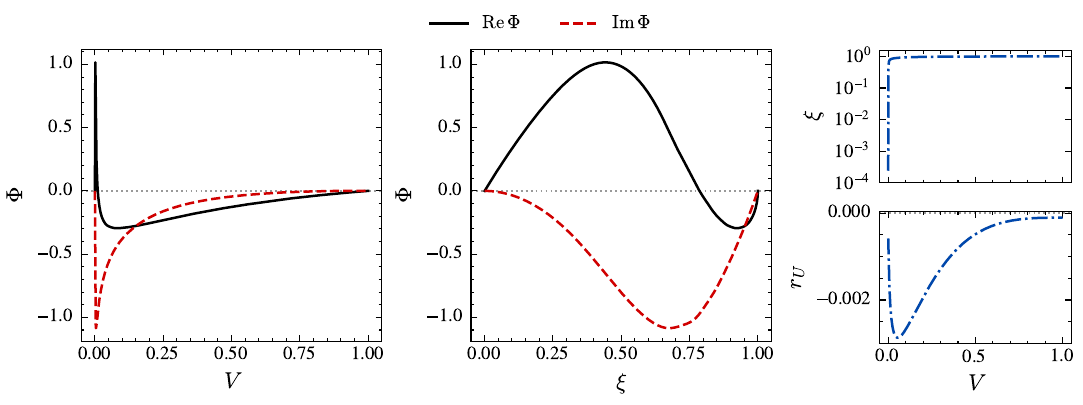}
	\end{center}
	\caption{The lowest-\(eM\) massless \(C^0\) gluing solution with \(eM=1.083\).\ \emph{Left}: scalar field profile as a function of \(V\);\ \emph{middle}: scalar field profile as a function of \(\xi\);\ \emph{right}: quantities \(\xi(V)\) and \(r_U(V)\) as functions of \(V\).}\label{fig:C0_fields_geometry}
\end{figure}

The lowest value of \(eM\) found was \(1.083\), a factor of \(8.6\) lower when compared to~\cite{gadioux2025formation}.
It was obtained with \(p=192\), instead of the \(p=1024\), which has the most degrees of freedom.
The admissibility constraints were satisfied with \(\inf_V\xi=2.41\times 10^{-4}\) and \(\sup_V r_U=-1.04\times 10^{-4}\). The scalar field profile is plotted in Figure~\ref{fig:C0_fields_geometry}, alongside \(\xi(V)\) and \(r_U(V)\).

Many observations aligned with those in~\cite{gadioux2025formation}.
Observe from the left of Figure~\ref{fig:C0_fields_geometry} that the scalar field profile is sharply peaked near \(V=0\).
This behaviour is supported by Equations~\eqref{eq:I_def} and~\eqref{eq:focusing_constraint}: both integrals contain quadratic dependence on \(\rho\), but the latter is weighted by \(V\), so that it is preferable for the amplitude to be focused at small \(V\).
As we can see in the right of Figure~\ref{fig:C0_fields_geometry}, \(\xi(V)\) rose sharply close to the endpoint value of 1 at very small \(V\); this led the plot of \(\Phi(V)\) to be stretched and produce the middle panel of Figure~\ref{fig:C0_fields_geometry}, when plotted against the physical quantity \(\xi\).

\subsubsection{Subextremal and Superextremal Horizons}\label{sec:sub_sup_horizon}
\begin{figure}[!htbp]
	\centering
	\includegraphics[width=0.48\textwidth]{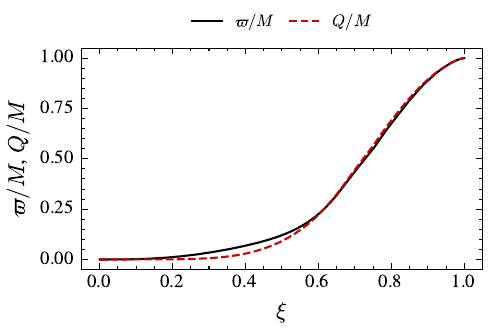}
	\caption{Charge profile \(Q / M\) and renormalised Hawking mass \(\varpi / M\) as functions of \(\xi\) for the lowest-\(eM\) massless \(C^0\) gluing solution with \(eM=1.083\).}\label{fig:best_roots_Q}
\end{figure}

Figure~\ref{fig:best_roots_Q} displays the charge \(Q / M\) and the renormalised Hawking mass \(\varpi / M\) as functions of \(\xi\) for the lowest-\(eM\) solution.
Observe that the charge transport is close to \(\varpi\), but there are subextremal and superextremal regions at small and large \(\xi\) respectively, as seen in~\cite{gadioux2025formation}.
Importantly, this behaviour is not a feature of the particular ansatz, but follows from the field equations and the gluing conditions at the endpoints. From the equations of motion and Equation~\eqref{eq:varpi},
\begin{equation}
  \varpi_V=\frac{Q}{r}Q_V+2r^2(-r_U)|\Phi_V|^2.
	\label{eq:Q-varpi_V}
\end{equation}
Through bounding arguments, we show that Equation~\eqref{eq:Q-varpi_V} can be integrated over sufficiently small intervals near the endpoints to find that the horizon is subextremal near the initial endpoint and superextremal immediately before the final extremal endpoint.\footnote{Charge and renormalised Hawking mass are plotted as functions of \( \xi \) for better visibility in Figure~\ref{fig:best_roots_Q}, but still the observation aligns with the analytical result in the \( V \)-coordinate since \( \xi(V) \) is strictly increasing (see Section~\ref{sec:amp_phase_decomposition} and the top right panel of Figure~\ref{fig:C0_fields_geometry}). This applies to Figure~\ref{fig:varpi_Q} and the discussions in Section~\ref{sec:hrmg} as well.} 
A detailed proof is given in Appendix~\ref{sec:sub_sup_proof}.

\subsubsection{Numerical Rigidity}
The LM continuation algorithm from Section~\ref{sec:continuation_algorithm} was initiated using the \(eM=1.083\) seed for \(C^0\) gluing, and it did not show any improvement and terminated at the same value of \(eM\). This further supports that the SLSQP search has found the lowest \(eM\) for this particular branch.
But one should note that the obtained value could be limited by the form of the ansatz.
The ansatz with the Type~2 (\(P_{2,0}\)) envelope (\(\delta_1=0.05\), \(\delta_2=0.1\)) was also tested with the same initial conditions, but it produced worse results (\(eM=1.092\), \(p=128\)) than the Type 1 (\(P_{1,0}\)) envelope.

\subsection{Higher-regularity Massless Gluing}\label{sec:hrmg}
For the higher-regularity gluing, we used the Type~2 envelope with \(\delta_1=0.05\) and \(\delta_2=0.1\) for the amplitude. The Type~2 envelope was chosen because we expect, from~\cite{gadioux2025formation} and the \(C^0\) gluing results (Section~\ref{sec:C0_res}), that the low-\(eM\) solutions are likely to have a sharp peak at small \(V\), while the Type~2 envelope allows the amplitude to grow rapidly at small \(V\), unlike the Type~1 envelope which becomes more suppressive at small \(V\) as the regularity \(k\) increases.

\begin{figure}[!htbp]
	\centering
	\includegraphics[width=0.98\textwidth]{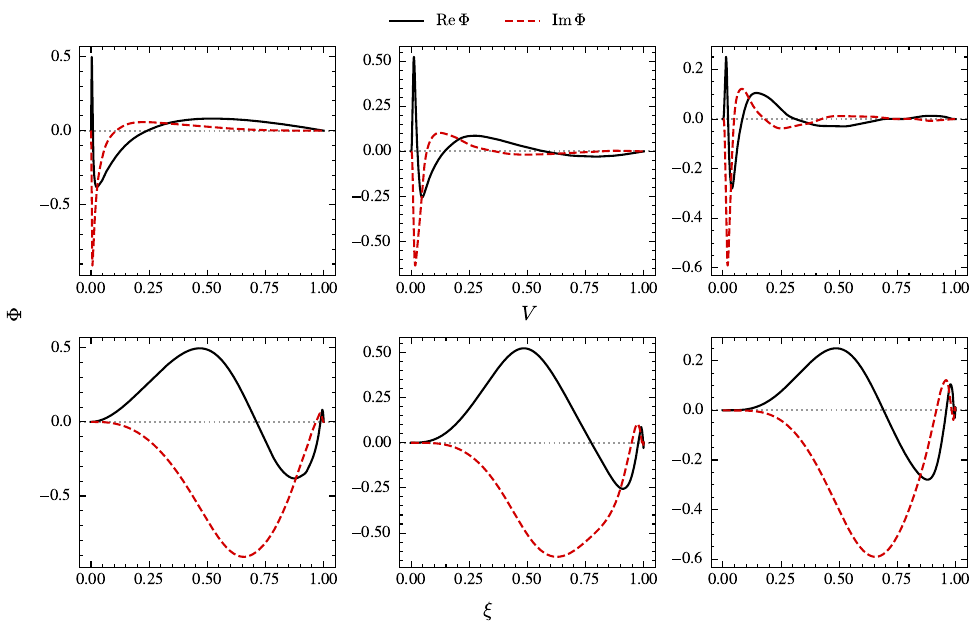}
	\caption{Scalar profiles \(\Phi\) of the lowest-\(eM\) massless gluing solutions plotted as a function of \(V\) in the first row and as a function of \(\xi\) in the second row.\ \emph{Left}: \(C^1\) gluing, \(eM=1.666\);\ \emph{middle}: \(C^2\) gluing, \(eM=2.735\);\ \emph{right}: \(C^3\) gluing, \(eM=3.353\).}
	\label{fig:Phi}
\end{figure}

\begin{figure}[!htbp]
	\centering
	\includegraphics[width=0.98\textwidth]{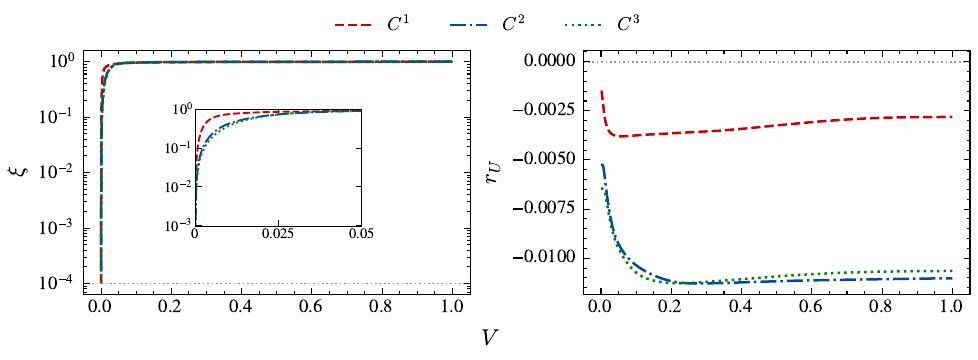}
	\caption{Plots of \(\xi(V)\) (\emph{left}) and \(r_U(V)\) (\emph{right}) of the lowest-\(eM\)  massless gluing solutions.}
	\label{fig:admissibility}
\end{figure}

\begin{figure}[!htbp]
	\centering
	\includegraphics[width=0.98\textwidth]{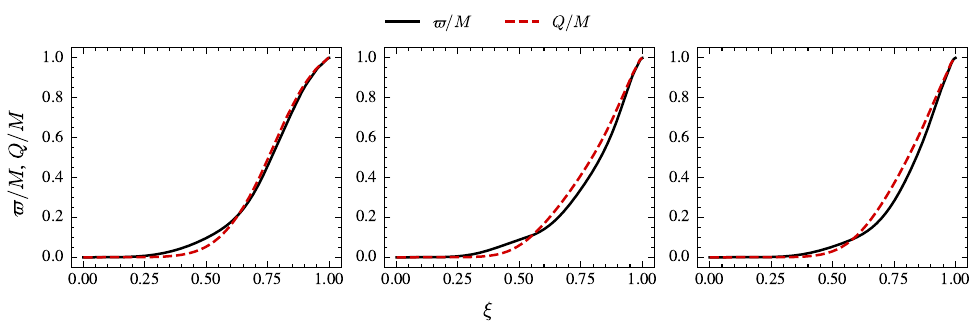}
	\caption{Charge \( Q / M \) and renormalised Hawking mass \( \varpi / M \) as functions of \( \xi \) for the lowest-\( eM \) gluing solutions.\ \emph{Left}: \( C^1 \); \emph{middle}: \( C^2 \);\ \emph{right}: \( C^3 \).}\label{fig:varpi_Q}
\end{figure}

We followed Section~\ref{sec:continuation_algorithm} and started decreasing \( eM \) from initial seeds with values \(eM=8, 20, 60\) and \(p=4,4,6\) for \(C^1\), \(C^2\) and \(C^3\) gluing respectively.\footnote{I thank Jorge E. Santos for providing a \( C^3 \) seed during the development of this work; the final computations used a different seed.}
The lowest values of \(eM\) obtained for \(C^1, C^2\) and \(C^3\) gluing are \(1.666, 2.735, 3.353\), with \(p=24, 2048, 48\) respectively.
These values improve the \( C^1 \) and \( C^2 \) results of~\cite{gadioux2025formation} by factors of approximately 47 and \(1.6\times 10^3\), respectively; a numerical \( C^3 \) result is obtained here for the first time.
The corresponding solutions are plotted in Figure~\ref{fig:Phi}.\footnote{As explained in Section~\ref{sec:redundant_dimensions}, for \( C^2 \) gluing, solutions with smaller \( p \) were used to plot Figures~\ref{fig:Phi},~\ref{fig:admissibility} and~\ref{fig:varpi_Q}.}

As noted in~\cite{gadioux2025formation}, many features of the \(C^0\) gluing solution are retained in the higher-regularity solutions.
From the top row of Figure~\ref{fig:Phi}, we immediately see that the scalar field profiles have sharp peaks at small \(V\). Again, this behaviour can be explained by the constraint argument in Section~\ref{sec:C0_res}.
Moreover, we observe that the rapid increase of \(\xi(V)\) at small \(V\) (left panel of Figure~\ref{fig:admissibility}) stretches \(\Phi\) when plotted against \(\xi\) (bottom row of Figure~\ref{fig:Phi}).
From these plots, we see that the locations of the dominant extrema of \(\Phi(\xi)\) are similar for all regularities, even though the amplitudes are different.
This is notable since the ansatz and the starting seed are different.
Quantitatively, \(\max_{\xi\in[0,1]}\Re\Phi (\xi)\) is attained at \( \xi=0.442,\, 0.467,\,0.485,\,0.486 \) and \( \min_{\xi\in[0,1]}\Im\Phi (\xi) \) is attained at \( \xi=0.675,\, 0.657,\,0.626,\,0.656\) respectively, for \(C^0,\,C^1,\,C^2,\,C^3\) gluing.
However, the quantities \( \Re\Phi \) and \( \Im\Phi \) are dependent on the choice of electromagnetic gauge. The gauge-independent amplitudes \( |\Phi| \) for the lowest-\( eM \) massless gluing solutions are plotted as functions of \( \xi \) in the left panel of Figure~\ref{fig:amplitude}, and we see that the peaks of \( |\Phi | \) do not coincide.
The number of oscillations increases with regularity: both \(\Re\Phi \) and \( \Im\Phi \) changed signs 1, 2, 3, 6 times for \(C^0,\,C^1,\,C^2,\,C^3\) gluing, respectively.\footnote{Some oscillations were too small to be visually identified in Figures~\ref{fig:Phi} and~\ref{fig:C0_fields_geometry}.}

\begin{figure}[!htbp]
	\begin{center}
		\includegraphics[width=0.98\textwidth]{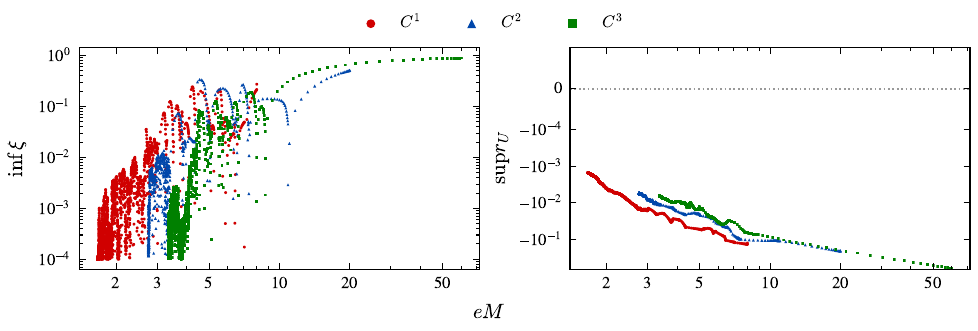}
	\end{center}
	\caption{Values of \( \inf_V\xi \) (\emph{left}) and \( \sup_Vr_U \) (\emph{right}) as functions of \(eM\) for the continuation run of the best \( C^1 \), \( C^2 \) and \( C^3 \) solutions.}\label{fig:margins_eM}
\end{figure}

As we can observe in Figure~\ref{fig:admissibility}, all of these solutions satisfied the admissibility conditions \(\inf_V\xi>0\) and \(\sup_V r_U<0\), but \(\inf_V\xi\) was the active constraint when trying to glue with lower \(eM\), as all solutions reached the imposed boundary of \(10^{-4}\) when running the continuation algorithm.
This behaviour is further illustrated in Figure~\ref{fig:margins_eM}. The value of \(\sup r_U\) showed a clear increasing trend as \(eM\) decreased, but it was safely away from the boundary for all solutions. Meanwhile, \(\inf\xi\) displayed a decreasing trend that reached the numerical limit \(\delta_\xi=10^{-4}\) multiple times as \(eM\) decreased, although there were some fluctuations in the trend, which is likely due to the margin rebuilding step in the continuation algorithm.
This suggests a natural experiment to see if decreasing the numerical buffer \(\delta_\xi\) would allow for a solution with lower \(eM\).
However, lowering the boundary further was not effective in decreasing \(eM\) further for our selected branches. The continuation algorithm was extended to \emph{lower} the boundary \(\delta_\xi\), trying values down to \(10^{-9}\), but the improvement in \(eM\) was below 0.01\% for both \(C^1\) and \(C^2\) gluing.

Charge \( Q /M  \) and renormalised Hawking mass \( \varpi /M \) are plotted as a function of \( \xi \) in Figure~\ref{fig:varpi_Q}.
As expected from the \(C^0\) analysis in Section~\ref{sec:sub_sup_horizon} and Appendix~\ref{sec:sub_sup_proof}, all three higher-regularity solutions exhibit the same qualitative behaviour: a subextremal region at small \(\xi\), followed by a superextremal region.
Across the \(C^1\), \(C^2\) and \(C^3\) solutions considered here, increasing regularity is accompanied by a larger maximum of \(Q-\varpi\) and a smaller maximum of \(\varpi-Q\). However, since the solutions also differ in their ansatz parameters, continuation branches and values of \(eM\), these trends cannot be attributed to regularity alone.
\subsubsection{Redundant Dimensions and Numerical Properties}\label{sec:redundant_dimensions}
\begin{figure}[!htbp]
	\begin{center}
		\includegraphics[width=0.98\textwidth]{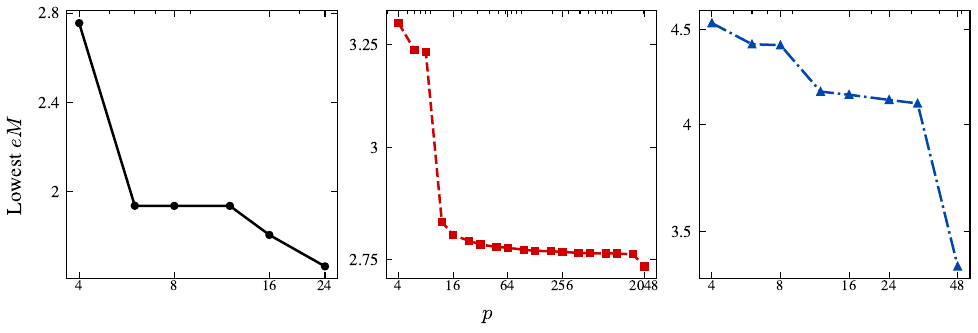}
	\end{center}
	\caption{The lowest \(eM\) values obtained for each value of \(p\) tested for \(C^1\) (\emph{left}), \(C^2\) (\emph{middle}) and \(C^3\) (\emph{right}) gluing solutions. }\label{fig:best_eM_by_p}
\end{figure}

Figure~\ref{fig:best_eM_by_p} displays the lowest \( eM \) obtained for each \( p \).
For \( C^2 \) gluing, the branch reached the maximum \( p=2048 \) set for the continuation algorithm, but increasing \( p \) was ineffective in decreasing \( eM \).

Nevertheless, the continuation algorithm was able to make meaningful progress at the largest \( p \) for all regularities.
We should note that the values obtained for the \(C^1\) and \(C^2\) gluing are not the lowest values achievable for the ansatz: for \(C^1\) and \(C^2\) gluing, although extremely slow, the continuation algorithm was still proceeding at the point of termination, with the average step size of \(\Delta(eM)=2.1\times 10^{-5},\,1.8\times 10^{-4}\) respectively for the last 50 successful steps. Unlike these, the continuation algorithm for \(C^3\) gluing was exhausted after trying all the recovery methods, and checking that increase in \(p\) was not effective in decreasing \(eM\) further.

These observations suggest that \( p=2048 \) is highly overparametrised for the \( \tanh \) ansatz, and lower values of \( p \) are sufficient to represent the relevant solution space. 
Motivated by this, we progressively reduced the network width by retaining the terms in Equation~\eqref{eq:tanh-ansatz} that contributed most strongly to \( \rho  \), \( \rho' \) for the amplitude or \( \omega \), \( \omega' \) for the phase.  
The remaining coefficients are fitted using linear least squares to approximate the original profiles and their derivatives. The resulting reduced ansatz was used as the initial seed for the constrained LM solver at fixed \(eM\) to match the residual. The reduced solution was retained only if it passed the same independent validation criteria (Appendix~\ref{sec:numerical_details}). 

The \( C^2 \) gluing solution was found with the same \( eM=2.735 \) but with \( p=48 \). Reduction of \( p \) was attempted for other regularities as well, but it could not be decreased further. Solutions with fewer parameters were used to plot Figures~\ref{fig:Phi},~\ref{fig:admissibility},~\ref{fig:varpi_Q} and~\ref{fig:amplitude}.\footnote{The scalar field profiles and all other physical quantities remained unchanged to the reported numerical precision.}

\subsection{Massive Scalar Field}\label{sec:massive_scalar_field}
A scalar field with a non-zero mass parameter was considered for \( C^1 \), \( C^2 \) and \( C^3 \) gluing. Following Section~\ref{sec:continuation_algorithm}, \( m /e \) was slowly increased while allowing \( eM \) to vary as well.
The best solutions reached \((m/e, eM)=(0.9944,46.80),\,(0.9961,29.99),\,(0.9866, 59.99)\) for \( C^1,\,C^2,\,C^3 \) gluing respectively. The Type~1 envelope was used for the \( C^1 \) and \( C^2 \) solutions, and the Type~2 envelope was used for the \( C^3 \) solution.

Reall has proved that an eRN black hole cannot be formed in a finite time from a gravitational collapse if \( m /e\geq 1\), for sufficiently regular solutions~\cite{reall2024third}. Thus, for sufficiently regular Minkowski-to-eRN gluing, \(m/e=1\) provides a theoretical upper threshold. Notably, all three values of \(m/e\) found here approach this threshold from below, even though the lower-regularity solutions considered in this work are not necessarily covered by the assumptions of~\cite{reall2024third}.

\begin{figure}[!htbp]
	\centering
	\includegraphics[width=0.98\textwidth]{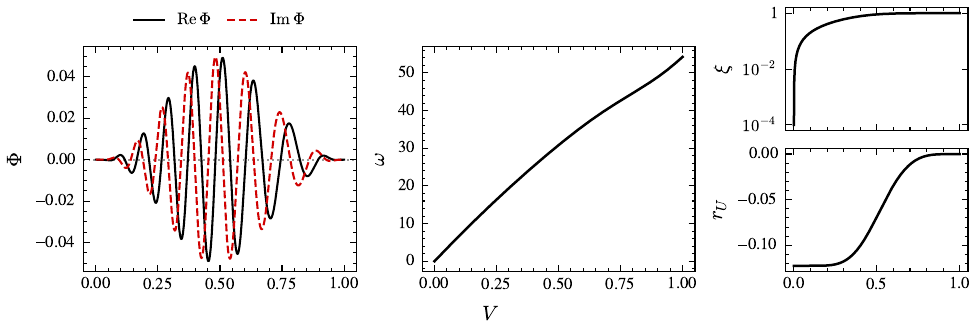}
	\caption{Largest-\(m/e\) massive \(C^2\) solution at \(m/e=0.9961\), \(eM=29.99\).\ \emph{Left}: scalar field;\ \emph{middle}: phase \(\omega\);\ \emph{right}: quantities \(\xi\) and \(r_U\).}\label{fig:massive_c2}
\end{figure}

\begin{figure}[!htbp]
	\centering
	\includegraphics[width=0.98\textwidth]{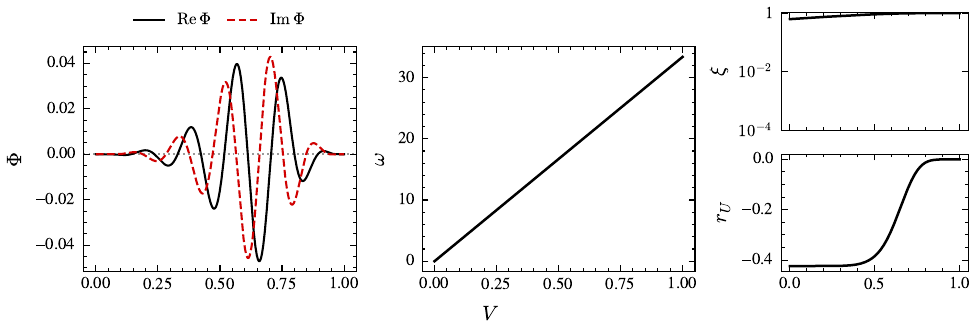}
	\caption{Largest-\(m/e\) massive \(C^3\) solution at \(m/e=0.9866\), \(eM=59.99\).\ \emph{Left}: scalar field;\ \emph{middle}: phase \(\omega\);\ \emph{right}: quantities \(\xi\) and \(r_U\).}\label{fig:massive_c3}
\end{figure}

Unexpectedly, the \(C^2\) gluing solution is closer to unity than the \( C^1 \) solution, while it is also a valid \(C^1\) gluing solution.\footnote{It is worth noting that the ansätze for different regularities are distinct and incompatible in this work (Equation~\eqref{eq:tanh-ansatz}) and they cannot be used as a seed for the other regularity, while every higher-regularity solution is automatically a lower-regularity solution as well.}
Hence, we will mainly focus on \( C^{2}  \) and \( C^3 \) solutions. The value \(m/e=0.9961\) is larger by a factor of \(9.031\) and \(100.4\) than the previous best results for \(C^1\) and \(C^2\) gluing in~\cite{gadioux2025formation}, respectively, and \( C^3 \) was not considered before.
Figures~\ref{fig:massive_c2} and~\ref{fig:massive_c3} display the real and imaginary parts of the scalar field, the phase \( \omega \) and the values of \( \xi\), \( r_U \) as functions of \( V \) for the highest-\( m /e \) gluing solutions found for \( C^2 \) and \( C^3 \) regularity, respectively.
Moreover, the amplitudes of the scalar fields of those solutions are plotted as functions of \( \xi \) in the right panel of Figure~\ref{fig:amplitude}.

From Figures~\ref{fig:massive_c2} and~\ref{fig:massive_c3}, we see that the phase \( \omega (V) \) is almost linear and reaches large values (greater than \( 10\pi \)), resulting in highly oscillatory profiles. Oscillatory behaviour is familiar for massive scalar test fields on black hole spacetimes. For example, Koyama and Tomimatsu found oscillatory late-time tails for massive scalar perturbations of nearly extremal Reissner--Nordström spacetime~\cite{koyama2001asymptotic, koyama2002slowly}.
However, we note that the setting is different, as we solve a fine-tuned nonlinear characteristic gluing problem with prescribed endpoint data, rather than the generic evolution of a test field on a fixed black hole background.

Also, unlike in the small-\(eM\) massless gluing solutions (Figure~\ref{fig:Phi}), the amplitude of the scalar field is no longer concentrated at early times, but is instead more broadly distributed. This aligns with the previous observation regarding the constraint given in Equation~\eqref{eq:focusing_constraint}: \(eM\) is large enough that there is more freedom for the amplitude to be distributed across \(V\) without violating the admissibility condition \(\inf_V\xi>0\).

Unlike \( \inf_V\xi \), the admissibility condition \( \sup_V r_U(V) \) was saturated for all solutions, with the values differing from the boundary \(- \delta_{r_U}=-10^{-4} \) by only \( 3.4\times 10^{-10} \), \( 5.2\times 10^{-10} \) and \( 6.7\times 10^{-9} \) for \(C^1\), \( C^2 \) and \(C^3\) gluing respectively (see also the right panels of Figures~\ref{fig:massive_c2},~\ref{fig:massive_c3}). This is expected from Equation~\eqref{eq:rr_U}: the positive term involving \( m^2 \) increases \( r_U \) and makes it more difficult to satisfy the admissibility condition \( \sup_V r_U(V)<0 \).

The final values were obtained with \(p=64\), \(p=4\) and \( p=4 \) for \(C^1\), \(C^2\) and \( C^3 \) gluing respectively.
That is, the \(C^1\) branch terminated at values of lower \(m/e\) despite having more trainable parameters and fewer residuals to satisfy. This suggests that the choice of branch is a more decisive factor than the regularity in this particular continuation problem.
Recall that the \(m/e\) continuation algorithm allowed the \(eM\) to vary in the direction extrapolated from the previous two steps. Note that the values of \(eM\) for all three solutions are clearly distinct, while the values of \(m/e\) are very close. This could be due to the different envelopes associated with the three regularities, but it is also possible that \(eM\) is not a decisive factor in increasing \(m/e\) in this continuation problem.

\section{Discussion}\label{sec:discussion}
Using the flexible \(\tanh\) ansatz, we found the lowest values of \(eM\) reached numerically for \(C^0\), \(C^1\), \(C^2\) and \(C^3\) gluing, and the largest values of \(m/e\) reached numerically for \(C^1\), \(C^2\) and \(C^3\) gluing, summarised in Table~\ref{tab:comparison-lambda-zero}.
\begin{table}[!htbp]
	\centering
	\caption{Best numerical results for asymptotically flat gluing to an extremal Reissner--Nordström black hole in~\cite{gadioux2025formation} and the present work.\ The massless entries give the minimum validated \(eM\), whereas the massive entries give the maximum validated \(m/e\). The regularity labels refer to the regularity imposed in each search; higher-regularity solutions also provide lower-regularity examples.}
	\label{tab:comparison-lambda-zero}
	\begin{tabular}{@{}lccc@{}}
		\toprule
		Optimised quantity            & Regularity & Ref.~\cite{gadioux2025formation} & This work    \\
		\midrule
		\multirow{4}{*}{\(\min eM\)}  & \(C^0\)    & \(9.27\)                         & \(1.083\)    \\
		                              & \(C^1\)    & \(78.5\)                         & \(1.666\)    \\
		                              & \(C^2\)    & \(4417\)                         & \(2.735\)    \\
		                              & \(C^3\)    & None                             & \(3.353\)    \\
		\midrule
		\multirow{3}{*}{\(\max m/e\)} & \(C^1\)    & \(0.1103\)                       & \(0.9944\)   \\
		                              & \(C^2\)    & \(0.00992\)                      & \(0.9961\)   \\
		                              & \( C^3 \)  & None                             & \( 0.9866 \) \\
		\bottomrule
	\end{tabular}
\end{table}
The lower bounds for \(eM\) improved substantially compared to~\cite{gadioux2025formation} for \(C^0\), \(C^1\) and \(C^2\) gluing. The differences between the lowest \(eM\) values found at different regularities were of the same order of magnitude, unlike~\cite{gadioux2025formation} (Table~\ref{tab:comparison-lambda-zero}).
In addition, a valid gluing solution was found for \(C^3\) gluing, corresponding to a classical (\(C^2\)) spacetime solution, again with \(eM\) of the same order of magnitude.

Very recently, for the massless \(C^0\) gluing, Schneider has found a sequence of analytic solutions with \(eM\) approaching \(((3+\sqrt{33})/3)^{1 /2}\approx 1.707\), and also proved that the necessary condition for which gluing is possible is \(eM>1/3\)~\cite{schneider2026chargebound}. Notice that our numerical bound \(1.083\) lies between the analytical results.
It would be interesting to find a new sequence of analytic solutions with \(eM\) close to \(1.083\), perhaps motivated by the early-peaked scalar profile observed. Furthermore, the current numerical result for massless \( C^0 \) gluing suggests that the bound \(eM>1/3\) is not sharp. Hence, a more ambitious work would be to improve this necessary condition.

The improvements when compared to~\cite{gadioux2025formation} were more substantial in the massive case. While the ansätze considered in the previous work were only able to provide solutions for the cases with \(m/e\ll 1\), we were able to get close to the upper threshold \(m/e=1\)~\cite{reall2024third} for all regularities considered. These results provide strong numerical evidence that the upper bound \(m/e<1\) is sharp for the \( C^k \) (\( k\leq 3 \)) gluing, and make it natural to conjecture that the bound remains sharp at arbitrary regularity. In Figures~\ref{fig:massive_c2} and~\ref{fig:massive_c3}, it was possible to observe a clear oscillatory behaviour inside an envelope for the scalar field profile. This suggests searching for an exact analytic solution of a similar shape, possibly a sequence of solutions of arbitrary regularity reaching the limiting value \(m/e=1\) from below with a sinusoidal profile.

Also, we have discussed in Section~\ref{sec:massive_scalar_field} that in addition to the regularity, the value of \(eM\) may not be important in the \(m/e\)-increase continuation problem.
It would therefore be interesting to study the dependence of the minimum \(eM\) on \(m /e\), in particular its behaviour as \(m/e\) approaches unity.

Although less dramatic than in~\cite{gadioux2025formation}, the lowest \(eM\) found (strictly) increased with the regularity of the solution.
This is expected numerically since the higher-regularity gluing conditions have more residuals to match (Equation~\eqref{eq:residual}), while the admissibility condition imposes the same constraint (Equation~\eqref{eq:focusing_constraint}).
However, it is unclear whether this behaviour is analytically true with a regularity-dependent sharp bound that follows from the equations of motion, boundary and admissibility conditions, or merely reflects the limitations of the ansatz and numerical algorithm.

As mentioned in Section~\ref{sec:continuation_algorithm}, different branches have displayed very distinct behaviours; multiple branches terminated at much higher \( eM \) values than those in Table~\ref{tab:comparison-lambda-zero}, for example.
A better understanding of the solution space may help to address this problem.
Moreover, one may try a different ansatz using basis functions rather than \(\tanh\), or a different envelope that is better suited to the nature of the profile, or to optimise perturbations to the current ansatz. It would also be interesting to explore whether the increase in the number of oscillations with regularity (Section~\ref{sec:hrmg}) is a numerical artefact or reflects a more general feature of the gluing solutions.

A straightforward extension of this work would be to consider the Schwarzschild to eRN gluing problem by modifying the initial condition for the Hawking mass. Furthermore, de Sitter/anti-de Sitter to eRN-(A)dS gluing with a non-zero cosmological constant could also be considered, as in~\cite{gadioux2025formation}, but with the improved ansatz and continuation algorithm.

In this work we have only constructed the solution on the null hypersurface \( \mathcal{C} \). It would therefore be very interesting to numerically extend the gluing solutions to construct spacetime solutions. In particular, we may obtain a \( C^2 \) spacetime solution from the \( C^3 \) gluing solution found.
Such evolutions could also provide a numerical setting in which to investigate the stability question discussed by Dafermos~\cite{dafermos2025stability}: whether the finite-time formation of an extremal black hole requires any additional degree of fine-tuning compared with the formation of a subextremal Reissner--Nordström black hole with prescribed mass-to-charge ratio.
\section*{Acknowledgments}
I am grateful to Maxime Gadioux and John R. V. Crump for valuable discussions, and to Jorge E. Santos for helpful discussions and for verifying some solutions.
I am particularly grateful to Harvey S. Reall for supervising this project and for many helpful discussions.
I am supported by a Trinity College Summer Studentship.
\section*{Data Availability}
The parameter values and other run data used in the paper are available in the GitHub repository: \href{https://github.com/junwoolee352/rn-characteristic-gluing}{GitHub}.
\printbibliography[heading=bibintoc,title={References}]
\appendix
\section{Higher Order Gluing Conditions}\label{sec:higher_order_gluing}
Explicit integrated expressions required for the higher order gluing conditions can be obtained from the Raychaudhuri, Maxwell and Wave equations (Equations~\eqref{eq:raychaudhuri}-\eqref{eq:Phi_UV}). 
For brevity, define
\[
L:=\log\Omega^2,\qquad
G:=-1+\frac{Q^2}{r^2}+m^2r^2|\Phi|^2,\qquad
S_\pm:=m^2r\pm\frac{ieQ}{r},
\]
\[
K_\pm:=\Omega^2S_\pm,\qquad
b:=ie(A_U)_V=-\frac{ie\Omega^2Q}{2r^2}.
\]
All quantities are evaluated on \(\mathcal C\), where \(\Omega^2=1\),
after taking transverse derivatives. Here \(D_U^j(\Phi_V)\) denotes
the covariant \(U\)-derivatives of \(\Phi_V\).

For \(C^2\) gluing,
\begin{align*}
r_{UU}(V)
&=\frac{1}{r(V)}
\left\{
r_U(0)^2+\frac14\int_0^V[L_UG+G_U](v)\,dv-r_U(V)^2
\right\},\\
D_U^2\Phi(V)
&=\frac{1}{r(V)}\Bigg\{
\int_0^V\Big[
-r_{UU}\Phi_V-r_UD_U(\Phi_V)\\
&\qquad
-\frac14\big((K_+)_U\Phi+S_+D_U\Phi\big)+rbD_U\Phi
\Big](v)\,dv-r_U(V)D_U\Phi(V)
\Bigg\}.
\end{align*}

For \(C^3\) gluing,
\begin{align*}
r_{UUU}(V)
&=\frac{1}{r(V)}\Bigg\{
\frac14\int_0^V
\big[(L_U^2+L_{UU})G+2L_UG_U+G_{UU}\big](v)\,dv\\
&\qquad-3r_U(V)r_{UU}(V)
\Bigg\},\\
D_U^3\Phi(V)
&=\frac{1}{r(V)}\int_0^V\Big[
-3r_UD_U^2(\Phi_V)-3r_{UU}D_U(\Phi_V)
-r_{UUV}D_U\Phi\\
&\qquad
-2r_{UV}D_U^2\Phi-r_{UUU}\Phi_V
-\frac14\big((K_- )_{UU}\Phi+2(K_-)_UD_U\Phi+S_-D_U^2\Phi\big)\\
&\qquad
+r\big(3bD_U^2\Phi+3b_UD_U\Phi+b_{UU}\Phi\big)
\Big](v)\,dv.
\end{align*}

The gluing residual is written in terms of ordinary \(U\)-derivatives.
Since \(\Phi(1)=0\),
\[
\Phi_U(1)=D_U\Phi(1),
\]
\[
\Phi_{UU}(1)
=D_U^2\Phi(1)-2ieA_U(1)D_U\Phi(1),
\]
and
\[
\Phi_{UUU}(1)
=D_U^3\Phi(1)-3ieA_U(1)D_U^2\Phi(1)
+3\left[(ieA_U(1))^2-ie(A_U)_U(1)\right]D_U\Phi(1).
\]

\section{Numerical Details}\label{sec:numerical_details}
\subsection{Initial Sampling}
For the randomly generated seeds, the parameters in Equation~\eqref{eq:theta} are sampled from several Gaussian and uniform distributions, with the precise scales and ranges varying between seeds. For example, in one commonly used amplitude initialisation, \(a_j\sim N(0,0.35^2)\) and \(w_j\sim U(4,12)\), while the remaining \( b_j \) are written as \(b_j=-w_jc_j\), with \(c_j\) sampled uniformly over a prescribed interval. Several values of \(\log\kappa\) were also tried manually. 
Seed discovery was first attempted with \( p=4 \), increasing to \(p=6\) when necessary.

\subsection{The Validation Process}\label{sec:the_validation_process}
The validation of the solution is done by independently evaluating the residual \(F_k(\theta)\) from the parameter \(\theta\) obtained from the main solver (SLSQP/LM), with the SciPy package~\cite{virtanen2020scipy}. The Raychaudhuri equation is solved using the eighth-order DOP853 integrator, and the remaining transport equations are evaluated using cumulative Simpson quadrature. Three different schemes are used when interpolating \(E(V)\) when integrating the Raychaudhuri equation: cubic interpolation on grids of 20001 and 40001 points, and PCHIP~\cite{fritsch1984method} interpolation on a grid of 40001 points. For a solution to be validated, we require \(\|F_k(\theta)\|_\infty<10^{-8}\) and \(\inf_V\xi>10^{-4}\), \(\sup_V r_U<-10^{-4}\) for all three schemes; furthermore we require consistencies between the three schemes, i.e.\ the largest change in the matching residual must be less than \(10^{-8}\) and the largest change in \(\inf_V\xi\) and \(\sup_V r_U\) must be less than \(10^{-5}\).

\subsection{The Continuation Algorithm}
The constrained LM algorithm is implemented using the PyTorch package and the associated automatic differentiation capabilities~\cite{paszke2019pytorch}. Inside the constrained LM algorithm, Equation~\eqref{eq:xi_raychaudhuri} is numerically solved on a uniform grid using a fourth-order Runge--Kutta scheme, implemented as a \( 2\times 2 \) transfer-matrix recurrence (which is much faster than the DOP853 integrator), while the remaining transport equations are evaluated in their integrated form using cumulative Simpson quadrature, which is also a fourth-order accurate scheme.
The parameters are optimised multiple times with successively finer grids (2001, 4001, 8001 and 16001 grid points) and only accepted if the largest component of the residual is less than \(10^{-8}\) at the finest grid, i.e. \(\|F_k(\theta)\|_\infty<10^{-8}\) and the admissibility conditions are satisfied with a margin of \(\delta_\xi=\delta_{r_U}=10^{-4}\).
The resulting candidate is then independently validated as described in Section~\ref{sec:the_validation_process}. The continuation step size is dynamically adjusted according to the success of the correction, with failed steps reduced and rapidly converged steps allowed to increase.
\subsubsection[eM-decrease Continuation]{\( eM \)-decrease Continuation}\label{sec:eM_dec_cont}
The continuation parameter is \(\lambda\coloneq 1/eM\), and the initial step size is set as \(3\times 10^{-4}\) for the \( eM \)-decrease continuation. If a solution is not found at a step size of \( 3 \times 10 ^{-4}  \), we try the \emph{margin rebuilding} step.

The margin can be expressed as
\[\mu(\theta)=\min\{\inf_V\xi(V;\theta),-\sup_V r_U(V;\theta)\},\]
and it is often the limiting factor in finding a solution. This is addressed by moving along the branch while preserving \(F_k=0\) to first order. Note that the kernel \(N=\ker\nabla_\theta F_k\) is the set of directions in which the residual does not change to first order, and \(\nabla_\theta\mu \) is the steepest direction of the margin increase. Hence, we may proceed in the direction \(P_N\nabla_\theta\mu\) where \(P_N\) is the orthogonal projection operator onto \(N\).
On each step in this direction, we optimise the parameters using the LM algorithm with the same value of \(eM\), while dynamically adjusting the step size. Each step is accepted only if \(\|F_k\|_\infty<10^{-8}\) and \(\Delta\mu>10^{-10}\), and the overall rebuilding is continued until several stopping conditions are met, e.g.\ when the direction is no longer effective: \(\|P_N\nabla_\theta \mu\|<10^{-12}\).

If a margin-rebuilt solution fails to proceed, even smaller step sizes are tried, with the smallest step size set as \(1\times 10^{-6}\).

If all of the above fails, we increase \(p\), up to 2048.
It is also increased when the continuation step is successful, but is \emph{stagnant} with the median of three previous successful step sizes less than \((3\times 10^{-4})/4=7.5\times 10^{-5}\).

\subsubsection[m/e-increase Continuation]{\( m /e \)-increase Continuation}
The continuation parameter here is \( m /e \). 
As discussed in Section~\ref{sec:continuation_algorithm}, we allow \( eM \) to vary between steps as well, i.e.\ \( (\theta,\log(eM)) \) is being predicted and corrected. Margin rebuilding in Section~\ref{sec:eM_dec_cont} is done for every successful solution found. 
Once the step size reaches \( 5\times 10^{-5} \), \( m /e \) is temporarily allowed to vary and pseudo-arclength type continuation~\cite{allgower1990numerical} is tried on the parameters \( (\theta,\log(eM),m /e) \). The increase in \( p \) is tried when the step reaches \( 10^{-8} \).

\section{Proof of the Subextremality and Superextremality}\label{sec:sub_sup_proof}
In this section, we prove that \( |Q(V)|<\varpi(V) \) (subextremal) and \( |Q(1-V)|>\varpi(1-V)\) (superextremal) for sufficiently small \( V>0 \) for the massless gluing solution considered in this work, by integrating Equation~\eqref{eq:Q-varpi_V}. More generally, the argument applies to any \(\Phi\in H_0^1((0,1);\mathbb C)\) that is not identically
zero in any neighbourhood of either endpoint. 

Let \( \Phi \in H_0^1((0,1);\mathbb{C}) \) be such a solution. 
From the admissibility constraints and continuity, there exist positive constants \( C,D \) such that 
\(|e|r^2(1+|Q| /r)\leq C \) and \( 2r^2(-r_U)\geq D \) throughout \( [0,1] \).

Let \(\epsilon\in(0,1)\).
By Cauchy-Schwarz, for \( 0\leq V\leq\epsilon \), 
\[
|\Phi(V) |\leq \int _0^V|\Phi_V(s)|ds\leq \sqrt{V}\left(\int_0^V|\Phi_V(s)|^2ds\right)^{1 /2}\leq \sqrt{\epsilon }\left(\int _0^\epsilon |\Phi _V|^2dV\right)^{1 /2}. 
\] 
From the Maxwell equation (Equation~\eqref{eq:Q_V}), \( |Q_V|\leq|e|r^2|\Phi | |\Phi _V| \). Combining these and using Cauchy-Schwarz one more time,
\begin{align*}
\int_0^\epsilon \left(1+\frac{|Q|}{r}\right)|Q_V|dV&\leq C \int _0^\epsilon |\Phi | |\Phi _V|dV \leq C\sqrt{\epsilon } \left(\int _0^\epsilon |\Phi _V|^2dV\right)^{1 /2}\int _0^\epsilon |\Phi _V|dV\\&\leq C\epsilon \int _0^\epsilon |\Phi _V|^2dV.
\end{align*}
Hence, integrating Equation~\eqref{eq:Q-varpi_V} from \(\varpi(0)=0 \) and using \( |Q(\epsilon)|\leq \int _0^\epsilon|Q_V|dV \) gives 
\begin{align*}
|Q(\epsilon)|-\varpi(\epsilon)&\leq \int _0^\epsilon \left(1+\frac{|Q|}r\right)|Q_V|dV-\int_0^\epsilon 2r^2(-r_U)|\Phi _V|^2dV\\
&\leq-(D-\epsilon C)\int _0^\epsilon |\Phi _V|^2dV<0
\end{align*}
if we pick \( \epsilon < D /C \). The proof is similar for the case near \( V=1 \); we have
\[
\int_{1-\epsilon}^1 \left(1+\frac{|Q|}{r}\right)|Q_V|dV\leq C\epsilon \int_{1-\epsilon}^1 |\Phi _V|^2dV.
\] 
Integrating Equation~\eqref{eq:Q-varpi_V} backwards from \( \varpi(1)=M\) and using \( Q(1-\epsilon)\geq M-\int_{1-\epsilon}^1|Q_V|dV \), we obtain 
\[
  |Q(1-\epsilon)|-\varpi(1-\epsilon )\geq (D-C\epsilon)\int_{1-\epsilon}^1 |\Phi _V|^2dV>0
\] 
if we pick \( \epsilon<D /C\).
\section{Additional Figures}
\begin{figure}[!htbp]
  \centering
  \includegraphics[width=0.98\textwidth]{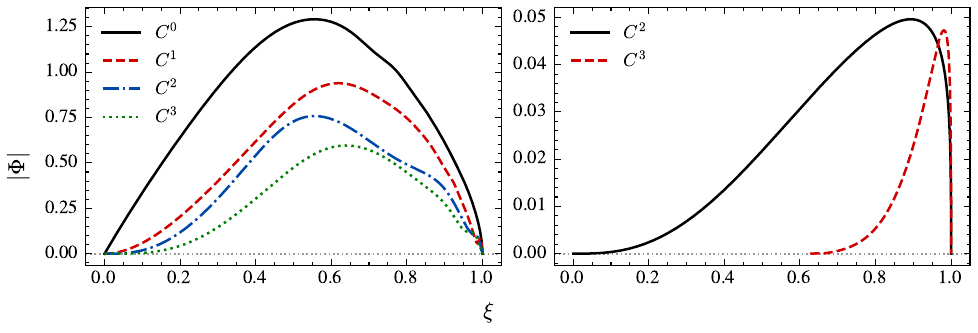}
\caption{Amplitude \( |\Phi| \) of a scalar field as functions of \( \xi\).\ \emph{Left}: lowest-\( eM \) massless \( C^0,\,C^1,\,C^2,\,C^3 \) gluing solutions.\ \emph{Right}: highest-\( m /e \) massive \( C^2,\,C^3 \) gluing solutions.}\label{fig:amplitude}
\end{figure}
\end{document}